\documentclass[12pt]{article}

\usepackage{geometry}
\usepackage{changepage}
\usepackage{bbm}
\usepackage[hidelinks]{hyperref}
\usepackage{amsmath,amsfonts,amssymb,amsthm, bm} 
\usepackage{graphicx} 
\usepackage{subcaption} 
\usepackage{adjustbox} 
\usepackage{float} 
\usepackage{algorithmicx} 
\usepackage{tabularx, comment} 

\newtheorem{theorem}{Theorem}

\newtheorem{lemma}{Lemma}

\floatname{algorithm}{Algorithm}
\usepackage{algorithm}
\usepackage{algpseudocode}

\title{Fast Structured Orthogonal Dictionary Learning using Householder Reflections}

\author{
  \begin{tabular}{cc}
    Anirudh Dash & Aditya Siripuram \\
    \small Department of Electrical Engineering & \small Department of Electrical Engineering \\
    \small Indian Institute of Technology, Hyderabad & \small Indian Institute of Technology, Hyderabad \\
    \small \texttt{ee21btech11002@iith.ac.in} & \small \texttt{staditya@ee.iith.ac.in}
  \end{tabular}
}
\date{}
\newcommand{\keywords}[1]{\textbf{Keywords:} #1}
\begin{document}

\maketitle

\begin{abstract}
    In this paper, we propose and investigate algorithms for the structured orthogonal dictionary learning problem. First, we investigate the case when the dictionary is a Householder matrix. We give sample complexity results and show theoretically guaranteed approximate recovery (in the \(l_\infty\) sense) with optimal computational complexity. We then attempt to generalize these techniques when the dictionary is a product of a few Householder matrices. We numerically validate these techniques in the sample-limited setting to show performance similar to or better than existing techniques while having much improved computational complexity.


\end{abstract}


\section{Introduction}
\label{sec:intro}

The orthogonal dictionary learning problem is posed as follows: Given a matrix \(\mathbf{Y} \in \mathbb{R}^{n \times p}\), can we find an orthogonal matrix \(\mathbf{V} \in \mathbb{R}^{n \times n}\) and a coefficient matrix \(\mathbf{X} \in \mathbb{R}^{n \times p}\) such that \(\mathbf{Y=VX}\)? Variants of this problem appear in standard sparse signal processing literature \cite{elad2010sparse} and signal processing-based graph learning approaches \cite{dong2019learning}, \cite{thanou2014learning}. Prior work in \cite{rusu2016fast, rusu2017learning} developed fast dictionary learning approaches assuming additional structure on the orthogonal matrix. The goal of this work is to build on this line of investigation to obtain recovery guarantees on \(\mathbf{V}\) and \(\mathbf{X}\) and sample complexity bounds under strong structural assumptions on the orthogonal matrix. We then attempt to extend some algebraic ideas from the solution to the case when some of the structural assumptions on $\mathbf{V}$ are relaxed.

The standard unstructured dictionary learning problem ($\mathbf{Y=DX}$) has been well investigated in literature. We refer to \cite{olshausen1997sparse, engan1999method, aharon2006k, mairal2009online, sun2015complete} as a few references.
The case when the dictionary is orthogonal is also well investigated: algorithms based on alternate minimization have been proposed \cite{lesage2005learning, bao2013fast}. Theoretical results pertinent to the above problem are usually of two kinds: proving the validity of proposed algorithms and identifying fundamental conditions (i.e., sample complexity or the number of columns $p$ required) for any algorithm to recover the factors $\mathbf{V}$ and $\mathbf{X}$.

This work focuses on the problem of orthogonal dictionary learning and is motivated by the following observations:
\begin{enumerate}
    \item Some applications, for example, graph learning, place additional structural assumptions on the orthogonal dictionary: for e.g. in graph learning, the orthogonal matrix \(\mathbf{V}\) is known to be an eigenvector matrix of a suitable graph.
    \item Even for unstructured orthogonal dictionary learning, attempts have been made to speed up the dictionary computation by approximating the dictionary as a structured orthogonal matrix \cite{rusu2016fast, rusu2017learning}. Most of the existing work is on unstructured orthogonal dictionary learning \cite{du2023matrix}, while work on structured orthogonal matrices in \cite{rusu2016fast, rusu2017learning} doesn't have sample complexity results.
    \item Most of the existing techniques are iterative and are sensitive to initialization \cite{rusu2016fast, rusu2017learning, liang2022simple}.
\end{enumerate}

We start the above investigation by assuming that the orthogonal matrix is a Householder matrix, similar to \cite{rusu2016fast}. We note that every orthogonal matrix can be expressed as a product of Householder matrices \cite{golub2013matrix, uhlig2001constructive}, thus allowing for the development of a new procedure to solve the orthogonal dictionary factorization problem. 


In this paper, we first analyze sample complexity for Householder matrices. By imposing a statistical model on the coefficient matrix $\mathbf{X}$, we show that recovery is possible with only \(\Omega(\log n)\) columns in \(\mathbf{Y}\) in the $l_{\infty}$ sense. The algorithm proposed utilizes the statistics of $\mathbf{Y}$ and is a non-iterative approach with theoretical guarantees for recovery. The computational complexity in learning the dictionary is $O(np)$, which is substantially smaller than previous methods such as \cite{rusu2016fast, zhai2020complete}. We then generalize these ideas to a product of multiple Householder matrices.

\section{Problem formulation and result summary}
\label{sec:prob}
Consider the setup of the unstructured orthogonal dictionary learning problem \( \mathbf{Y = VX}\), where \(\mathbf{Y} \in \mathbb{R}^{n \times p} \) is the data matrix, \(\mathbf{X} \in \mathbb{R}^{n \times p} \) is an (unknown) sparse representation matrix and \(\mathbf{V = \mathbf{H}_1\mathbf{H}_2\ldots \mathbf{H}_m} \in \mathbb{R}^{n \times n} \) is a product of $m$ Householder matrices $\mathbf{H_1, H_2, \ldots, H_m}$: with \( \mathbf{H_i = I - 2u_iu_i^\textsf{T}}\), where \(\mathbf{u_i}\) are (unknown) unit-norm vectors. Given the data matrix \(\mathbf{Y}\), we want to estimate \(\mathbf{V}\) and \({\mathbf{X}}\).

We refer by \(u_i\) the entries of the vector \(\mathbf{u}\), and denote by $||\mathbf{u}||_\infty = \max |u_i|$, the infinity-norm of $\mathbf{u}$. We denote by $||\mathbf{A}||_F$ (with $||\mathbf{A}||^2_F = \text{Trace}(\mathbf{A^\textsf{T}A})$) the Frobenius norm of matrix $\mathbf{A}$. We refer by $X_{ij}$ the entries of matrix $\mathbf{X}$.


We use the following sparsity model on \(\mathbf{X}\): the support is drawn from an iid Bernoulli distribution with parameter \(\theta\):
\begin{equation}
    \label{eq:1*}
    \begin{aligned}
        X_{ij} \neq 0 \text{ w.p. } \theta, 0 \text{ w.p. } 1-\theta.
    \end{aligned}
\end{equation}

\noindent All entries on the support are drawn from an i.i.d Uniform distribution in the range $[1, 2]$ (See Section \ref{sec:pfs} for more details). Let $\mu$ be the mean of this distribution. We assume that $\theta$ and $\mu$ are known. 
We attempt to investigate the following questions in this work:
\begin{enumerate}
    \item How many columns $p$ in $\mathbf{Y}$ are required to estimate $\mathbf{V}$ to reasonable accuracy?
    \item How does the recovery of $\mathbf{V}$ depend on the sparsity $\theta$ in $\mathbf{X}$ and how robust is this recovery to errors in $\mathbf{Y}$ ?
    \item What is the computational complexity of this estimate?
\end{enumerate}

In section \ref{sec:structured}, we analyze the case when $m=1$. We show that with $p=\Omega(\log n)$\footnote{Note that we say \(f(n)= \Omega(g(n))\) if \(\lvert f(n) \rvert \geq  C \lvert g(n) \rvert \) for some constant \(C\) for all \(n\) large enough.} columns in $\mathbf{Y}$, it is possible to recover the underlying vector $\mathbf{u}$ accurately in the $l_\infty$ sense with $O(np)$ computations non-iteratively (as opposed to $O(n^2p)$ per iteration with a standard Procrustes based solution (see Section \ref{sec:prior})). Building on the ideas from Section \ref{sec:structured}, we propose algorithms for the case $m>1$ in Section \ref{sec:general}. We demonstrate with numerical experiments that the proposed algorithm improves both approximation error and computational performance compared to existing solutions when the number of columns $p$ is low.

We also use the following: if $\mathbf{H}$ is an $n\times n$ Householder matrix, computing $\mathbf{Hx}$ for a vector $\mathbf{x}$ costs $O(n)$ arithmetic operations, as opposed to $O(n^2)$ for an arbitrary matrix.

\section{Prior work}
\label{sec:prior}
The standard orthogonal dictionary learning problem is formulated as follows (given a dataset $\mathbf{Y} \in \mathbb{R}^{n \times p}$ and a fixed sparsity level of $s$): \( \text{arg min}_{\mathbf{V, X}}  \{\| \mathbf{Y} - \mathbf{VX} \|_F^2: \mathbf{V^\textsf{T} V} = \mathbf{I}, \quad \|\mathbf{x}_i \|_0 \leq s, 1 \leq i \leq n \}  \),
where $\|\mathbf{x_i}\|_0$ is the number of non-zero entries in the $i^\textsf{th}$ column of $\mathbf{X}$.

Solving this involves a standard alternating minimization solution \cite{lesage2005learning}: When $\mathbf{V}$ is fixed, the estimate $\mathbf{\hat{X}}$  is updated by thresholding the product $\mathbf{V^\textsf{T}Y}$. When $\mathbf{X}$ is fixed, the estimate $\mathbf{\hat{V}}$ is updated via Orthogonal Procrustes \cite{ten1977orthogonal} ($\mathbf{\hat{V} = UW^\textsf{T}}$, given the Singular Value Decomposition $\mathbf{YX^\textsf{T}}=\mathbf{U\Sigma W^\textsf{T}}$). These updates are done iteratively till convergence.

However, the technique above is computationally expensive due to an SVD (of $\mathbf{YX^\textsf{T}}$) in each iteration (thus costing $O(n^2 \text{max}\{n,p\})$ per iteration). Note that non-orthogonal dictionary learning techniques \cite{tropp2004greed}, \cite{tropp2004just}  have similar computational complexity while having better representational performance. Consequently, \cite{rusu2016fast, rusu2017learning} provide fast orthogonal transform learning techniques by assuming some additional structure on the orthogonal dictionary, leading to improved computational performance. The work in \cite{rusu2016fast} assumes the orthogonal dictionary to be a product of $O(\log n)$ Householder reflections and provides an iterative algorithm to estimate the dictionary. The work in \cite{rusu2017learning} generalizes this approach using Givens rotations.

This work builds on the prior work by investigating the sample complexity (number of columns required) and robustness under statistical assumptions on the sparse representation $\mathbf{X}$. We completely analyze the case when $m=1$ using concentration inequalities and propose algorithms for the general case that improves on prior work under these statistical assumptions in the sample limited case (i.e., $p<n$).
Due to the statistical assumptions, our approach also has the advantage of being non-iterative and non-spectral, as opposed to prior work.
\section{Recovery for the structured orthogonal (Householder) dictionary}
\label{sec:structured}
In this section, we give sample complexity results for the case when $\mathbf{Y=HX}$, the matrix $\mathbf{H}$ is Householder ($\mathbf{H= I - 2 uu^\textsf{T}}$) and the sparse representation $\mathbf{X}$ follows the statistical model described in \eqref{eq:1*}. We see that using the first-order statistical properties of the induced distribution on $\mathbf{Y}$ is sufficient to estimate $\mathbf{H}$ to high accuracy.

\begin{theorem}
    \label{thm:1}
    (\emph{Householder Recovery}) Consider $\mathbf{Y=HX}$ and the model described in \eqref{eq:1*} for $\mathbf{X}$. Suppose
    \begin{enumerate}
        \item the unit vector \(\mathbf{u}\) defining the Householder matrix \(\mathbf{H}\) satisfies
              $c = \sum u_i = \Omega(n^\alpha)$ for $\alpha>1/4$,
              \item $u_ic$ is bounded for all $i$, and
        \item the number of samples/columns $p> C\log n / \theta^2 \mu^2$, with constant $C$ large enough;
    \end{enumerate}

    then \(\mathbf{u}\) can be recovered (up to sign) with the following recovery guarantee:
    \[
        \mathbb{P} \left ( \lVert \mathbf{u} - \hat{\mathbf{u}} \rVert_\infty > t\right ) \leq O\left((1/n)^{t^2 O(1)}\right).
    \]

    We get
    \(\mathbb{P} \left ( \lVert \mathbf{u} - \hat{\mathbf{u}} \rVert_\infty > t\right ) \rightarrow 0. \)
    The estimate $ \hat{\mathbf{u}} $ is computed via Algorithm \ref{find_H_X}, and the computational complexity involved in calculating $ \mathbf{\hat{u}} $ is \(O(np)\).
\end{theorem}

Sample complexity reduces with an increase in $\theta$ since we are able to extract more information from the data matrix as fewer entries in $\mathbf{X}$ are $0$. The entries of the Householder matrix multiplying it are fixed, and this is leveraged when $\mathbf{X}$ has a larger support.


\begin{algorithm}[H]
    \caption{Finding $\mathbf{H, \ X}$ for $\mathbf{Y=HX}$}
    \label{find_H_X}
    \textbf{Input:} $\mathbf{Y}, \theta, \mu$ \\
    \textbf{Output:} $\mathbf{H, \ X}$
    \begin{algorithmic}[1]
        \State Set $c^2 = \left(n-\sum_{i=1}^{n}\sum_{j=1}^{p} {Y_{ij}}/p\theta \mu\right)/2 $, with $c\geq 0$
        \State set $u_i= \left(1-\sum_{j=1}^{p}
            {Y_{ij}}/p\theta \mu\right)/2c$, for $i=1:n$
        \State Set $\mathbf{H} = \mathbf{I-2uu^\textsf{T}}$
        \State Set $\mathbf{X'} = \mathbf{H}^\textsf{T}\mathbf{Y}$
        \State Set $\mathbf{X} =$ $HT_{\zeta}$($\mathbf{X'}$)
    \end{algorithmic}
\end{algorithm}

Note that once $\mathbf{H}$ is obtained, we estimate the sparse representation $\mathbf{X}$ by computing $\mathbf{HY}(=\mathbf{H^{\textsf{T}}Y})$ and thresholding entry-wise. Note that this operation can be performed in $O(np)$ arithmetic operations (since $\mathbf{HY = Y - 2 u u^\textsf{T} Y}$ can be computed in $O(np)$). Thus both $\mathbf{H}$ and $\mathbf{X}$ are estimated in $O(np)$.
We provide numerical simulations for this algorithm in Section \ref{sec:sim}.

\noindent \textbf{Remark:} \(HT_{\zeta}(\cdot)\) is the hard threshold operator (i.e., \( HT_{\zeta}(x) = x\cdot \mathbb{I}(|x| \geq \zeta) \)). The value $\zeta$ is chosen heuristically. Furthermore, we are implicitly using the following: if $\mathbf{u}$ is a solution, then $-\mathbf{u}$ is also a solution, as both produce the same Householder matrix.

\section{Recovery for the general orthogonal dictionary}
\label{sec:general}

We move to the case when the orthogonal dictionary $\mathbf{V = H_1H_2\ldots H_m}$ is a product of $m>1$ Householder reflectors. Prior work in \cite{rusu2016fast} proposes an alternating iterative technique to estimate the Householder matrices. Following up from Section \ref{sec:structured}, we propose a sequential update strategy to estimate $\mathbf{H_i}$. Unlike the earlier work, this update is non-iterative (involves a fixed $m$ number of steps). Before discussing the algorithm, we note the following fundamental limitation to recovering the Householder matrices $\mathbf{H_i}$ in this setup.

\begin{lemma}
    \label{lem1}
    For any Householder matrix $\mathbf{H}$, there exist Householder matrices $\mathbf{H_1, H_2, H_3}$ different from $\mathbf{H}$ such that $\mathbf{HH_1 = H_2H_3}$.
\end{lemma}
Since the Householder matrices cannot be uniquely identified, we consider the error metric in this setup as $||\mathbf{V-\hat{V}}||_F$ (as opposed to an error in the individual $\mathbf{u_i}$). 

We first note the following modification of Algorithm \ref{find_H_X}.
Suppose $\mathbf{Q}$ is an orthogonal matrix, and let $\mathbf{Y = HQX}$ where $\mathbf{X}$ satisfies the statistical model from \eqref{eq:1*}. If $\mathbf{Q}$ is known, then $\mathbf{H}$ can be recovered from $\mathbf{Y}$ in a very similar fashion to the approach described in the proof of Theorem \ref{thm:1}. We skip the steps due to space constraints, and refer to Section \ref{sec:alg-H-Q-X} for additional details. The modified algorithm is summarized in Algorithm \ref{find_H_X2}.
\begin{algorithm}[!ht]
    \caption{Finding $\mathbf{H, \ X}$ for $\mathbf{Y=HQX}$}
    \label{find_H_X2}
    \textbf{Input:} $\mathbf{Y, Q}, \theta, \mu$ \\
    \textbf{Output:} $\mathbf{H, \ X}$
    \begin{algorithmic}[1]
        \State set $s_i= \sum_{j=1}^{n}{Q_{ij}}$, for $i=1:n$
        \State set $k_i= \left(s_i-\sum_{j=1}^{p}
            {Y_{ij}}/p\theta \mu\right)/2$ for $i=1:n$
        \State set $u_i= k_i/ \left(\sum_{m=1}^{n}
            k_ms_m\right)^{1/2}$ for $i=1:n$
        \State Set $\mathbf{H} = \mathbf{I-2uu^\textsf{T}}$
        \State Set $\mathbf{X'} = (\mathbf{HQ})^\textsf{T}\mathbf{Y}$
        \State Set $\mathbf{X} =$ $HT_{\zeta}$($\mathbf{X'}$)
    \end{algorithmic}
\end{algorithm}

Following this, we use a sequential strategy to update the $\mathbf{H_i}$. We consider a general case initialization to elaborate upon our algorithm. In step $i$, we note that
\[
    Y = \underbrace{\mathbf{H_1H_2\ldots H_{i-1}}}_W\mathbf{H_i}\ \underbrace{\mathbf{H_{i+1}H_{i+2}\ldots H_m}}_Q \mathbf{X},
\]
so that $\mathbf{W^\textsf{T}Y = H_i Q X}$. Applying Algorithm \ref{find_H_X2} to $\mathbf{W^\textsf{T}Y}$ gives us $\mathbf{H_i}$. We do this for $i=1,2,\ldots,m$ to obtain estimates for $\mathbf{H_i}$, and then set the estimate $\mathbf{\hat{V}}$ as the product of the obtained estimates for $\mathbf{H_i}$. This is outlined in Algorithm \ref{find_V_X}.
Computationally, the algorithm requires updating the data matrix $\mathbf{Y}$ for each $i=1:m$, and each update requires multiplying with a Householder matrix, costing $O(np)$ for each $i$. Note that we only need the sum of the entries of $\mathbf{Q}$ for each row; the sequential precomputation of these vectors $\mathbf{Q}\mathbf{1}$ for all $i=1:m$ costs $O(mn)$. Finally, Algorithm \ref{find_H_X2} costs $O(np)$ and must be repeated at each update step. Thus, the overall computational complexity of Algorithm \ref{find_V_X} is $O(nmp)$.

The spectral technique \cite{rusu2016fast} involves computing $\mathbf{XY^\textsf{T}}$ (which costs $O(n^2p)$) followed by finding the eigenvectors of an $n\times n$ matrix for each step (i.e. $m$ times). This process is repeated for a certain number of iterations till convergence. In contrast, the proposed approach is non-iterative and costs $O(npm)$ (i.e., the complexity scales linearly with $n)$. This reduction is achieved due to statistical assumptions on the sparse representation $\mathbf{X}$.

\begin{algorithm}[H]
    \caption{Finding $\mathbf{V}$ for $\mathbf{Y=VX}$}
    \label{find_V_X}
    \textbf{Input:} $\mathbf{Y}, m, \theta, \mu$ \\
    \textbf{Output:} $\mathbf{V}$
    \begin{algorithmic}[1]
        \State Initialize $\mathbf{H_i}$
        \State Set $Z[m+1] = \mathbf{1}$
        \State For $i=m$ to $1$, set $Z[i]=\mathbf{H_{i}}Z[i+1]$
        \For{For $i=1$ to $m$}
        \State Find $\mathbf{H_i}$ using Algorithm 2, using $Z[i]$ as $\mathbf{Q}\mathbf{1}$
        \State Update $\mathbf{Y}$ as $\mathbf{H_i^\textsf{T}Y}$
        \EndFor
        \State Set $\mathbf{V=\prod_i H_i}$
    \end{algorithmic}

\end{algorithm}

\section{Simulations}
\label{sec:sim}
In this section, we show some numerical results on the approximation error and robustness of the proposed algorithm.
\subsection{Data Generation}
The ground truth Householder matrices \(\mathbf{H_i = I - 2u_iu_i^\textsf{T}}\) are generated by selecting each entry of \(\mathbf{u}_i\) randomly (i.i.d Gaussian/Uniform) and then normalizing the obtained vector. The support of \(\mathbf{X} \) is generated by using i.i.d Bernoulli entries for each entry \({X_{ij}}\) with parameter \(\theta\). The non-zero entries are filled using i.i.d Uniform samples in the range $[1,2]$. With $\mathbf{H_i}$ and $\mathbf{X}$, the data matrix $\mathbf{Y}$ is computed as $\mathbf{Y = H_1H_2\ldots H_m X + N}$ where $\mathbf{N}$ is a noise matrix with i.i.d zero mean Gaussian entries. In most experiments, the number of rows $n$ is set to \(n=1000\), and the number of columns $p$ varies from \(2\) to $18$.

\subsection{Results for the case \texorpdfstring{$m=1$}{m=1}}
In Fig \ref{fig:3}, we plot the \(l_{\infty}\) error in \(\mathbf{u}\) (on the $y$-axis) with varying number of columns in \(\mathbf{Y}\); for different sparsity regimes. As we can see, the error decreases with an increase in the number of columns, as expected. The error when \(\theta\) is lower is slightly higher than the corresponding error for larger values of \(\theta \), which is consistent with Theorem \ref{thm:1}. Figure \ref{fig:4} shows the average per entry error in (Frobenius norm sense) \(\mathbf{X}\), with varying number of columns. Finally, in Figure \ref{fig:5}, we plot the estimation error under different SNR\footnote{The Signal-to-Noise Ratio (SNR) in decibels (dB) is given by: \(\text{SNR}_{\text{dB}} = 10 \log_{10} \left({P_{\text{signal}}}/{P_{\text{noise}}} \right) \) where \( P_{\text{signal}} \) and \( P_{\text{noise}} \) represent the signal and noise power, respectively.} regimes. As is evident from the results, the algorithm is relatively robust to noise.

\subsection{ Results for the general case}

Next, we provide results for orthogonal matrix recovery in a sample-limited setup. The orthogonal matrix \(\mathbf{V}\) is generated as a product of \(m\) Householder matrices \(\mathbf{H_1, H_2, \cdots H_m}\), (where the Householder vector \(\mathbf{u_i}, \ i \in \{1,2,\cdots m\}\) is generated by choosing each entry randomly and then normalizing the vector) i.e., \(\mathbf{V=H_1 H_2 \cdots H_m}\).

\begin{figure}[htbp]
    \centering
    \includegraphics[width=0.4\textwidth]{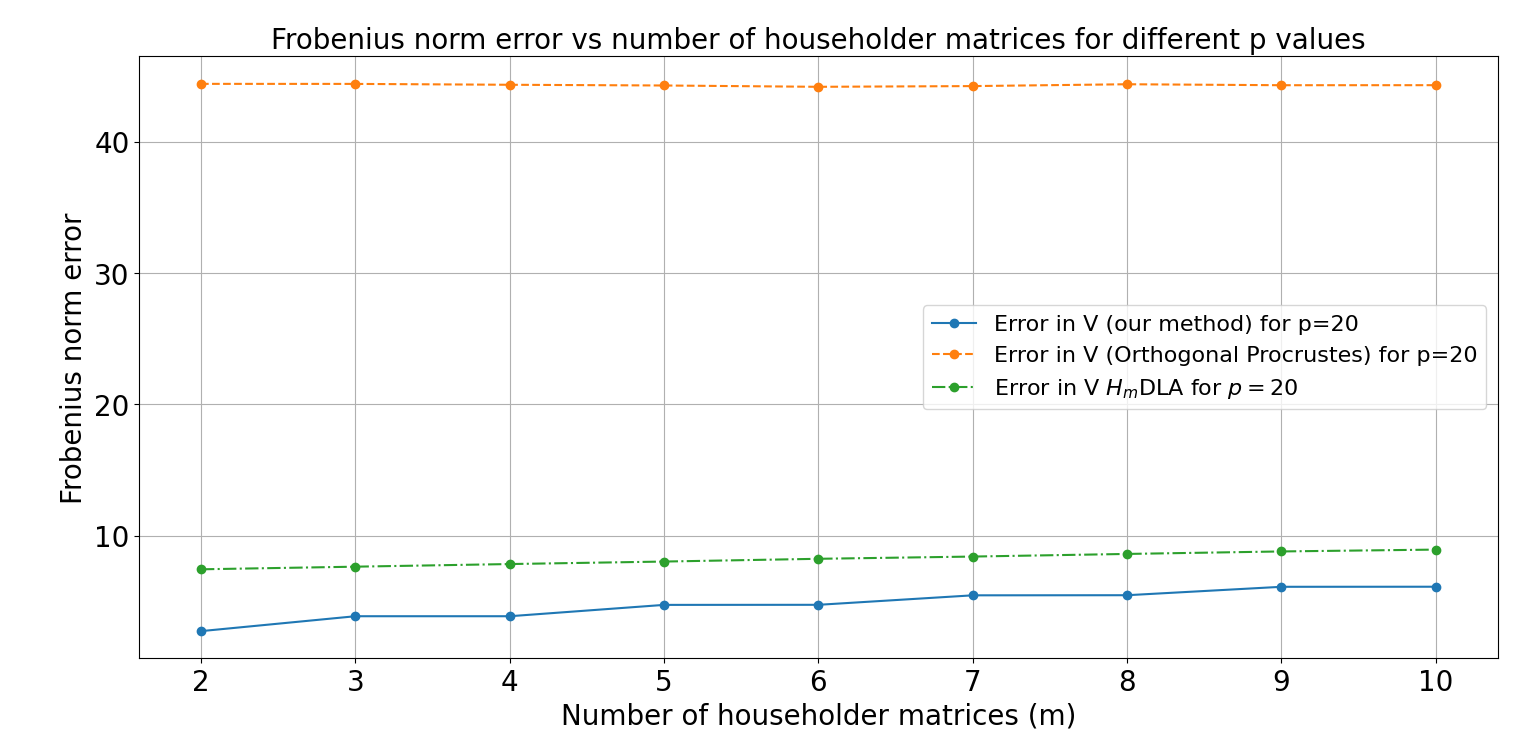} %
    \hfill
    \caption{Frobenius norm error in the estimated orthogonal dictionary for a varying number of Householder matrices (n=1000; p=20)}
    \label{fig:1}
\end{figure}

\begin{figure}[htbp]
    \centering
    \includegraphics[width=0.4\textwidth]{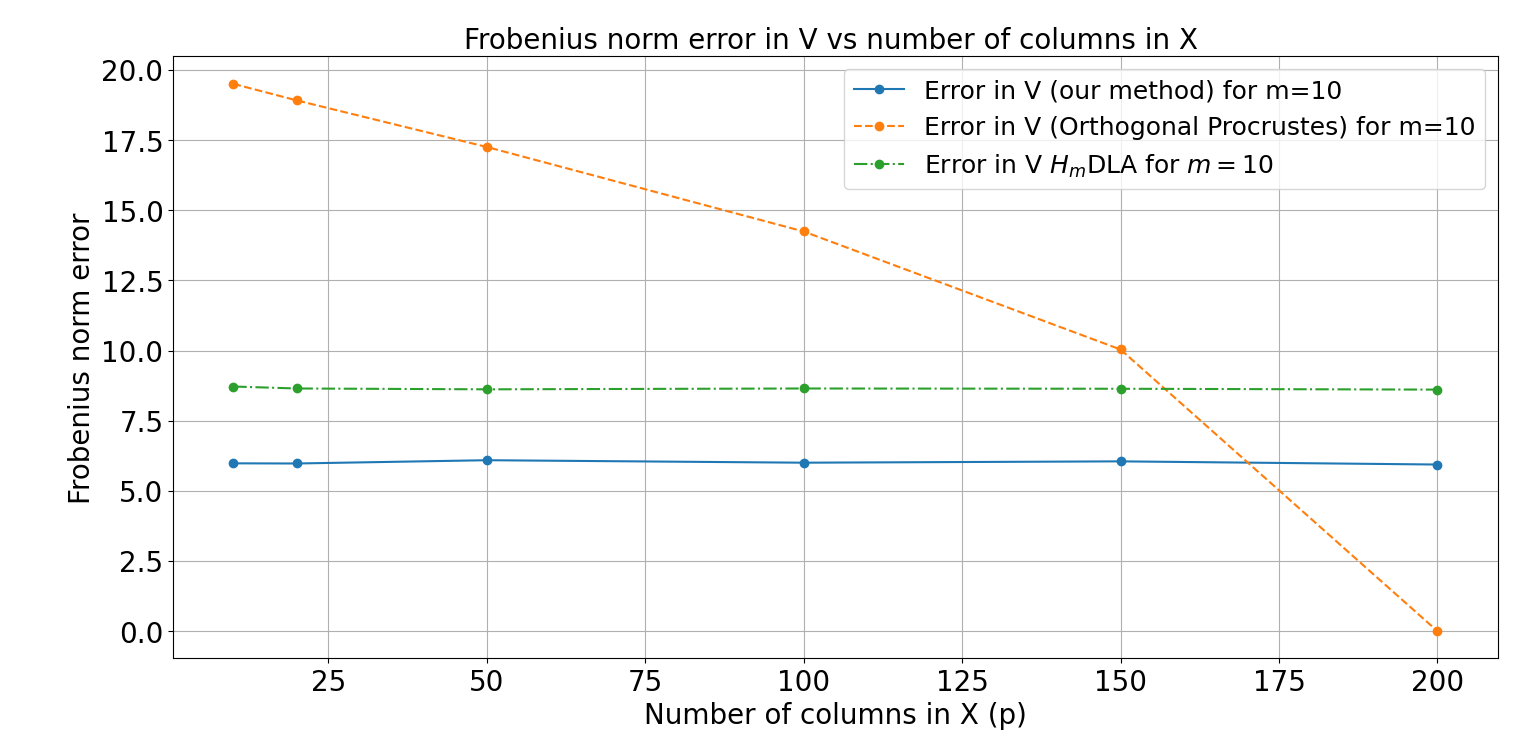} %
    \hfill
    \caption{Frobenius norm error in the estimated orthogonal dictionary for a varying number of columns $(n=1000; m=10)$}
    \label{fig:2}
\end{figure}

\begin{figure}[htbp]
    \centering
    \includegraphics[width=0.4\textwidth]{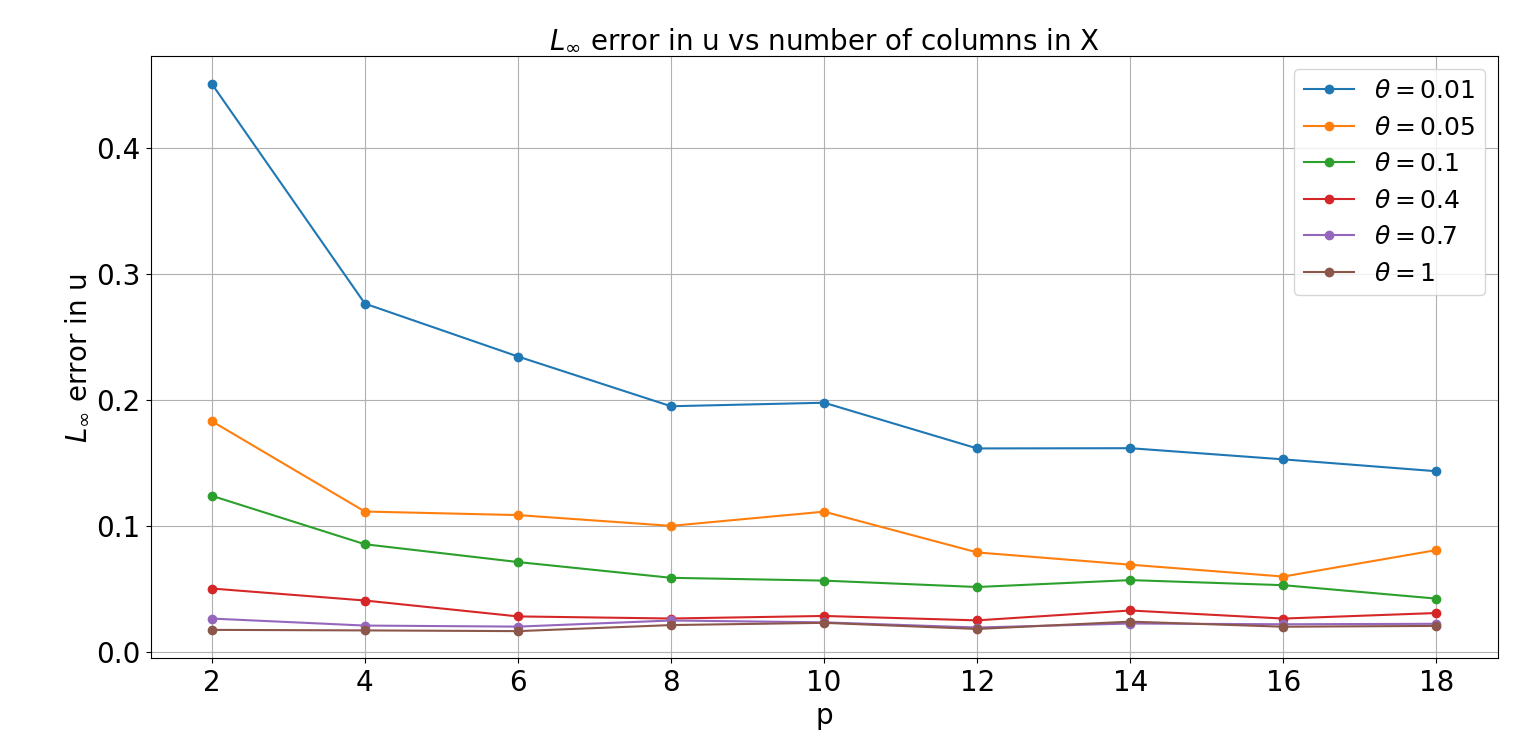} %
    \hfill
    \caption{$l_{\infty}$ norm error in $\mathbf{u}$ for a varying number of columns and different sparsity levels ($\theta$) for $\mathbf{Y=HX}$ $(n=1000)$}
    \label{fig:3}
\end{figure}

\begin{figure}[htbp]
    \centering
    \includegraphics[width=0.4\textwidth]{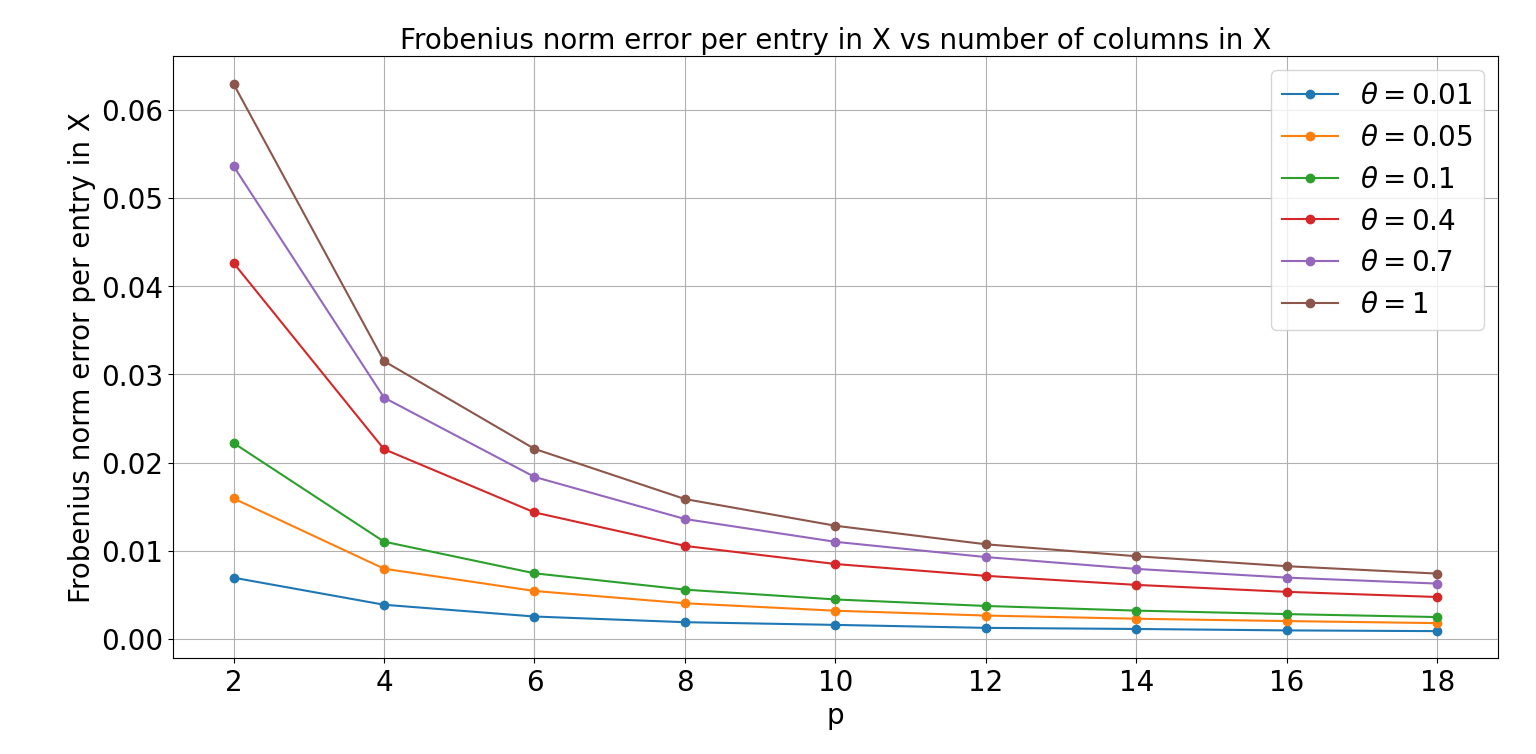} %
    \hfill
    \caption{Frobenius norm error per entry in $\mathbf{X}$ for a varying number of columns and varying sparsity levels ($\theta$) for $\mathbf{Y=HX}$ $(n=1000)$}
    \label{fig:4}
\end{figure}

\begin{figure}[htbp]
    \centering
    \includegraphics[width=0.4\textwidth]{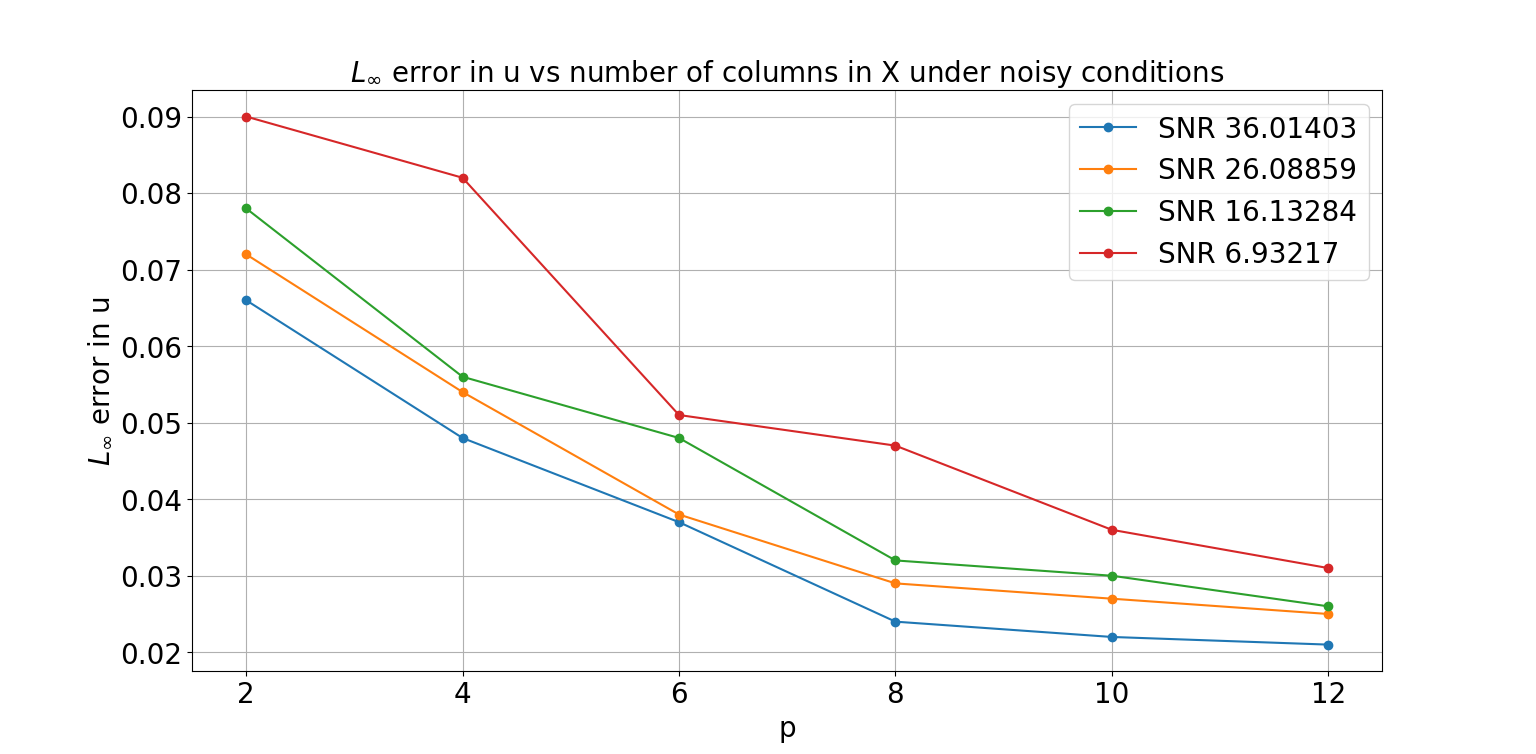}
    \hfill
    \caption{$l_{\infty}$ norm error in $\mathbf{u}$ for a varying number of columns and different sparsity levels ($\theta$) for $\mathbf{Y=HX}$ under noisy conditions (SNR is in dB) $(n=1000)$}
    \label{fig:5}
\end{figure}

We initialize $\mathbf{H_1, H_2, \ldots}$ to $\mathbf{I}$ for our experiments. 
The algorithm works well even if we initialize the matrices as arbitrary Householder matrices, since their deviation from the identity matrix is not too large, considering the multiple cases of choosing the vector $\mathbf{u}$.
In Fig \ref{fig:1}, we plot the Frobenius norm error in 
$\mathbf{V}$ for our method, the algorithm proposed in 
\cite{rusu2016fast}, and the solution to the Orthogonal Procrustes problem (for the Orthogonal Procrustes solution, we assume we know $\mathbf{X}$ exactly: so this is the best case performance of Procrustes) for a varying number of Householders, with $p=20$. In Fig \ref{fig:2}, we repeat the above experiment for varying columns for a fixed $m=10$. In this case, we used $n=200$, as opposed to the other experiments, since we increased $p$ to a relatively larger value of $200$. As the plots show, our method performs significantly better than the best-case Procrustes solution (i.e., with known $\mathbf{X}$) when we have very few samples and does slightly better than the method proposed in \cite{rusu2016fast} but with much better computational complexity.

Note that the figure shown is for the case when the number of Householder matrices is $m=10$. 
The theoretical results proved are for a single Householder case $m=1$ only. The error for this 
$m=1$ case decreases with an increase in the number of columns, as expected. The plot in Fig 2 
is for a heuristic algorithm for a product of multiple Householders ($m=10$). We suspect that 
our assumption on the constituent $\mathbf{u}$ vectors being orthogonal is causing some numerical 
errors to propagate as $m$ increases. Empirically, 
we observed that the usefulness of our algorithm lies in the sample-limited setting, where
Procrustes performs much worse. 

\section{Proofs and other details}
\label{sec:pfs}

\subsection{Proof of Theorem \ref{thm:1}}
Given that $c = \sum u_i = \Omega(n^{1/4})$ we assume without loss of generality that $c>0$. First, we note that due to the Householder structure of the matrix $\mathbf{H}$, we have \( {Y_{ij}} =\sum_{k=1}^{n} {H_{ik}} {X_{kj}} = \sum_{k=1}^{n} (\delta_{ik}-2u_iu_k) {X_{kj}}\). Taking expectations,
\begin{equation}
    \mathbb{E}[ {Y_{ij}} ]=\theta \mu \sum_{k=1}^{n} ( \delta_{ik} - 2u_iu_k) = \theta \mu(1-2u_ic),
    \label{eq:EY-c-u-relation}
\end{equation}

so that \( \sum_{i=1}^{n}\sum_{j=1}^{p}\mathbb{E}({Y_{ij}}) = \theta \mu(n-2c^2)p\). We estimate this expectation empirically, estimating $c = \sum u_i$ first, and consequently $u_i$. Following \eqref{eq:EY-c-u-relation}, the estimate for $c$ is
\[
    \hat{c}^2 = \left(1-\frac{\sum_{i=1}^{n} \sum_{j=1}^{p} {Y_{ij}}}{np\theta \mu}\right)\frac{n}{2}, \quad \hat{c}\geq 0.
\]
We note that $\sum_i \sum_j {Y_{ij}} =\sum_i \sum_j \sum_k {{H_{ik} X_{kj}}}$ is a weighted sum of independent random variables, so we use the Hoeffding's inequality \cite{hoeffding1994probability} to bound the error in the estimate of $c$:
\begin{equation}
    \mathbb{P}\left( \left\lvert c - \hat{c} \right\rvert > t \right) \leq \mathbb{P}\left( \left\lvert c^2 - \hat{c}^2 \right\rvert > t \right)    \leq 2 \exp\left( -8  t^2 \theta^2\mu^2 p \right),
    \label{eq:error-estimate-c}
\end{equation}
So  $\hat{c} = c + O(1)$ w.h.p\footnote{We say that an event $A_n$ holds with high probability (w.h.p) if $P(A_n) \geq 1-1/n^\alpha$ for some $\alpha >0$}. We skip the algebra due to space constraints. However, this follows from a direct application of Hoeffding's inequality.
Following the estimate of $c$ from above, using \eqref{eq:EY-c-u-relation} we estimate the entries of the ground truth vector $\mathbf{u}$ generating the Householder matrix as
\[
    \hat{u}_i = \frac{1}{2\hat{c}}\left(1-{\sum_{j=1}^{p} {Y_{ij}}}/{p\theta \mu} \right) := \frac{\hat{y}}{2\hat{c}}.
\]

Note that the error in the estimate \(\hat{u}_i\) has two components - due to the error in $\hat{c}$ and the deviation in the empirical mean above. 
Reapplying the Hoeffding's inequality to $\sum_j {Y_{ij}} = \sum_j \sum_k {H_{ik}X_{kj}}$, we obtain the error in the empirical mean as
\begin{equation}
    \label{eq:error-estimate-ui-1}
    \mathbb{P}\left[\left \lvert \hat{y}/2c - u_i\right \rvert\geq t  \right] \leq 2 \exp\left(-8t^2c^2\theta^2 \mu^2 p\right).
\end{equation}
Now we have
\begin{align*}
     & \mathbb{P}\left[\left \lvert \hat{u}_i - u_i\right \rvert \geq t  \right]=\mathbb{P}\left[\left \lvert \hat{y}/2\hat{c} - u_i\right \rvert \geq t  \right]                  \\
     & \quad \leq \mathbb{P}\left[\left \lvert \hat{y}/2\hat{c} - \hat{y}/2{c} \rvert \geq t/2 \right] + \mathbb{P} \left[\lvert \hat{y}/2{c} - u_i\right \rvert\geq t/2  \right], \\
     & \quad \leq \mathbb{P} \left[  \lvert \hat{y}(c-\hat{c})/c\hat{c} \rvert \geq t  \right] + 2 \exp \left(-2t^2c^2\theta^2 \mu^2 p\right).
\end{align*}
Note that from \eqref{eq:error-estimate-ui-1}, $\hat{y} = 2u_ic + O(1) $ w.h.p, and from \eqref{eq:error-estimate-c} and the hypothesis of the theorem, $c\hat{c}$ is $\Omega(n^{1/2})$. Therefore $\hat{y}/c\hat{c} $ is $O(1)$ w.h.p, and a repeat application of \eqref{eq:error-estimate-c} gives us
\begin{align*}
     & \mathbb{P}\left[\left \lvert \hat{u}_i - u_i\right \rvert \geq t   \right]             \\
     & \quad \leq 4\exp(-8 \theta^2 \mu^2 t^2 O(1) p)(1-1/n^\alpha) \text{ for some }\alpha>0 \\
     & \quad \leq 4 \left(\frac{1}{n}\right)^{8Ct^2}\left(1- \frac{1}{n^\alpha}\right)
\end{align*}
By union bound over all the $u_i$'s, the theorem follows. Note, as long as the mean ($\mu$) is bounded away from zero, the above inequality continues to hold, and the proof generalizes. For example, if $\mu = O(\sqrt{1/n})$, the final inequality will not hold, and the number of columns $p$ required for the guaranteed recovery proposed would in fact be $\Omega(n\text{log}n)$ instead of $\Omega(\text{log}n)$ (from Theorem \ref{thm:1}). We picked $[1,2]$ just to simplify this part of the proof. Note that if we consider the support values to be uniform on $[a,b]$ (instead of $[1,2]$), Equation 3
changes to 
\begin{align*}
 \mathbb{P}\left( \left\lvert c - \hat{c} \right\rvert > t \right) \leq \mathbb{P}\left( \left\lvert c^2 - \hat{c}^2 \right\rvert > t \right)    \leq 2 \exp\left( -8  t^2 \theta^2\mu^2 p / (b-a)^2 \right),
\end{align*}
so as long as $b-a = \Omega(1)$, the theoretical results hold (i.e. no change in order for the number of columns $p$ required); only the values in the simulations will change slightly.  







We show an equivalent method to recover \(\mathbf{u}\). Define
\begin{equation}
    \label{eq:H-X-alt}
    k_i = \left(1- {\sum_{j = 1 }^{p} {Y_{ij}}}/{p\theta \mu} \right)/2
    = u_{i}\left( \sum_{z=1}^{n} u_z \right) = u_{i}c,
\end{equation}
From this, we have \({k_m}{u_i} = {k_i}{u_m}\). Thus, we have \(k_i^2 = u_{i}^2\left( \sum_{m=1}^{n} {k_m} \right) \). Using the unit norm property of $\mathbf{u}$, we obtain an estimate of $u_i$ using $k_1, k_2, \ldots$. The computational complexity involved in calculating these estimates (using either approach) is \(O(np)\). 


\subsection{Proof of Lemma \ref{lem1} }

Let the Householder vectors corresponding to the matrices \( \mathbf{H, H_1, H_2, H_3} \) be \( \mathbf{u, u_1, u_2, u_3} \), respectively. Given $\mathbf{u}$, we choose \( \mathbf{u_1} \) such that \( \mathbf{u_1} \) is orthogonal to \( \mathbf{u} \).  Then we have \[ \mathbf{HH_1} = (\mathbf{I} - 2 \mathbf{uu}^\top)(\mathbf{I} - 2 \mathbf{u_1 u_1}^\top) = \mathbf{I} - 2 \mathbf{uu}^\top - 2 \mathbf{u_1 u_1}^\top. \]
Now we set \( \mathbf{u_2} = (\mathbf{u} + \mathbf{u_1}) / \sqrt{2} \) and \( \mathbf{u_3} = (\mathbf{u} - \mathbf{u_1}) / \sqrt{2} \), so that $\mathbf{u}_1$ and $\mathbf{u}_2$ are unit norm and orthogonal. We can now verify with basic algebra that
\begin{align*}
    \mathbf{H_2 H_3} & = \mathbf{I} - 2 \mathbf{u_2 u_2}^\top - 2 \mathbf{u_3 u_3}^\top = \mathbf{I} - 2 \mathbf{u u}^\top - 2 \mathbf{u_1 u_1}^\top = \mathbf{H}\mathbf{H_1}. 
\end{align*}

\subsection{Details on Algorithm \ref{find_H_X2}}
\label{sec:alg-H-Q-X}
The key difference in Algorithm \ref{find_H_X2} is how the entries of the matrix $\mathbf{Q}$ enter the computation of the estimates. We set \(s_i=\sum_{k=1}^{n} {{Q_{ik}}}\) as the entries of $\mathbf{s}=\mathbf{Q}\mathbf{1}$, so that \eqref{eq:EY-c-u-relation} modifies to \(\mathbb{E}[ {Y_{ij}} ]= \theta \mu (s_i-2u_i \mathbf{u}^\textsf{T}\mathbf{s}) \). Accordingly, \eqref{eq:H-X-alt} modifies to $k_i=\left(s_i-\sum_{j=1}^{p} {Y_{ij}}/p\theta \mu\right)/2=u_i \mathbf{u}^\textsf{T}\mathbf{s}$. Thus, we have \(k_i^2 = u_{i}^2\left( \sum_{m=1}^{n} s_m {k_m} \right) \), from which we obtain an estimate of $u_i$ using $k_1, k_2, \ldots$.  This approach reduces to that of Algorithm \ref{find_H_X} when $\mathbf{Q=I}$ (since $s_i = 1$ for all $i$). Note that $u_i$ follows the sign of $k_i$ and $\sum_{m=1}^{n} k_ms_m \neq 0$ is a necessary condition for the above to work. The proof of the theoretical guarantee for this approach is very similar in structure to that described in Theorem \ref{thm:1}. 

\vspace{-0.2cm}

\end{document}